\documentclass[a4paper,fleqn]{cas-dc}

\usepackage[authoryear]{natbib}
\usepackage{amsmath,amsfonts}
\usepackage{algorithmic}
\usepackage{algorithm}
\usepackage{wasysym}

\def\tsc#1{\csdef{#1}{\textsc{\lowercase{#1}}\xspace}}
\tsc{WGM}
\tsc{QE}

\begin{document}
\let\WriteBookmarks\relax
\def\floatpagepagefraction{1}
\def\textpagefraction{.001}

\shorttitle{A Survey of PPG's Application in Authentication}    
\shortauthors{L. Li et al.} 

\title[mode = title]{A Survey of PPG's Application in Authentication}  



%

\author[1]{Lin Li}
[orcid=0000-0001-7497-9002]






\affiliation[1]{organization={Swinburne University of Technology},
            city={Melbourne},
            postcode={3000}, 
            state={VIC}, 
            country={Australia}}

\author[2]{Chao Chen}





\affiliation[2]{organization={RMIT University},
            city={Melbourne},
            postcode={3000}, 
            state={VIC},
            country={Australia}}



\author[3]{Lei Pan}




\affiliation[3]{organization={Deakin University},
            city={Waurn Ponds},
            postcode={3216}, 
            state={VIC},
            country={Australia}}

\author[4]{Leo Yu Zhang}



\affiliation[4]{organization={Griffith University},
            city={Gold Coast},
            postcode={4222}, 
            state={QLD},
            country={Australia}}

\author[5]{Zhifeng Wang}
\cormark[1]


\affiliation[5]{organization={Nantong First People's Hospital},
            city={Nantong},
            postcode={226006}, 
            state={Jiangsu},
            country={China}}

\author[1]{Jun Zhang}



\author[1]{Yang Xiang}




\begin{abstract}
Biometric authentication prospered because of its convenient use and security. 
Early generations of biometric mechanisms suffer from spoofing attacks.
Recently, unobservable physiological signals (e.g., Electroencephalogram, Photoplethysmogram, Electrocardiogram) as biometrics offer a potential remedy to this problem.
In particular, Photoplethysmogram (PPG) measures the change in blood flow of the human body by an optical method. 
Clinically, researchers commonly use PPG signals to obtain patients' blood oxygen saturation, heart rate, and other information to assist in diagnosing heart-related diseases.
Since PPG signals contain a wealth of individual cardiac information, researchers have begun to explore their potential in cyber security applications. 
The unique advantages (simple acquisition, difficult to steal, and live detection) of the PPG signal allow it to improve the security and usability of the authentication in various aspects.
However, the research on PPG-based authentication is still in its infancy. 
The lack of systematization hinders new research in this field.
We conduct a comprehensive study of PPG-based authentication and discuss these applications' limitations before pointing out future research directions.  
\end{abstract}



\begin{keywords}
Biometrics \sep Physiological Signals \sep Photoplethysmogram \sep User Authentication
\end{keywords}

\maketitle


\section{Introduction}
Authentication ensures the legitimacy of access to data \citep{WANG2020107118} and the identity of individuals. 
Authentication is useful in many areas of our lives, including commercial applications, healthcare, access control, and many more. 
There are three categories of authentication---knowledge-based authentication like passwords, object-based authentication, like ID cards, and biometric-based authentication, like face recognition \citep{jain2006biometrics}. 
Biometric-based authentication uses physiological or behavioral characteristics extracted from a person as a source of idiosyncratic information \citep{HuangPCRAuth}. 
It does not suffer from being forgotten compared with knowledge- and object-based methods.
Since each human has many idiosyncratically physical or behavioral characteristics, a wealth of individual information can be leveraged to strengthen biometric-based authentication against fabrication.
The traditional features used for biometrics include fingerprint, face, iris, voice, palmprint, and many more \citep{jia2021108122}. 
In the 2010s, biometric authentication thrived, for example, using face recognition to unlock a smartphone and fingerprint recognition to unlock a door. 
Nevertheless, these early versions of biometric authentication are often vulnerable to presentation attacks \citep{wang2020differences, Jascha9316670}. 
A presentation attack means that an attacker impersonates a legitimate user to present biometrics to an authentication system. 
A common scenario is using a 3D mask representing the victim's face to fool the face recognition system.

Physiological signals are considered as biometrics because they are not readily observable. 
Such signals include Electroencephalogram, Electrocardiogram, and Photoplethysmogram (PPG) \citep{WANG2020107381,9130771Huang,Hwang9130730}. 
Specifically, a Pentagon's product uses infrared lasers to detect people's unique heart features to authenticate individuals \citep{hambling2019pentagon}; 
a Canadian company Nymi has developed an authentication system using wrist-worn pulse sensors as an alternative to fingerprint recognition \citep{eberz2017broken}.
Different from traditional biological features, physiological signal-based features are invisible on the human body's skin surface, making it challenging to be collected and analyzed by attackers from remote locations.

Among the physiological signals, PPG is a non-invasive optical method for measuring the volume of light absorbed or reflected by microvascular in biological tissues \citep{Keerthana9447911}.
Furthermore, PPG has a wide range of research prospects in authentication due to its unique advantages: 
\textbf{Simple acquisition}---An oximeter or a camera alone can capture PPG signals from a human body.
Furthermore, PPG sensors embedded in wearable devices simplify and reduce the cost of PPG signal acquisition.  
\textbf{Difficult to steal}---Traditional biometrics are subject to many easy attacks.
Fingerprints and palmprints can be extracted from touchscreen surfaces left by a user \citep{vachon2020identity}, while facial images can be taken at a distance.
In contrast, contact-based PPG signals are not directly exposed to the attacker, making them difficult to spoof.
\textbf{Live detection}---The liveliness of the users involved in the system is ensured by the natural liveness detection system because the PPG signal responds to the information of the human heartbeat. 

\begin{figure*}[!ht]
  \centering
  \includegraphics[width=\linewidth]{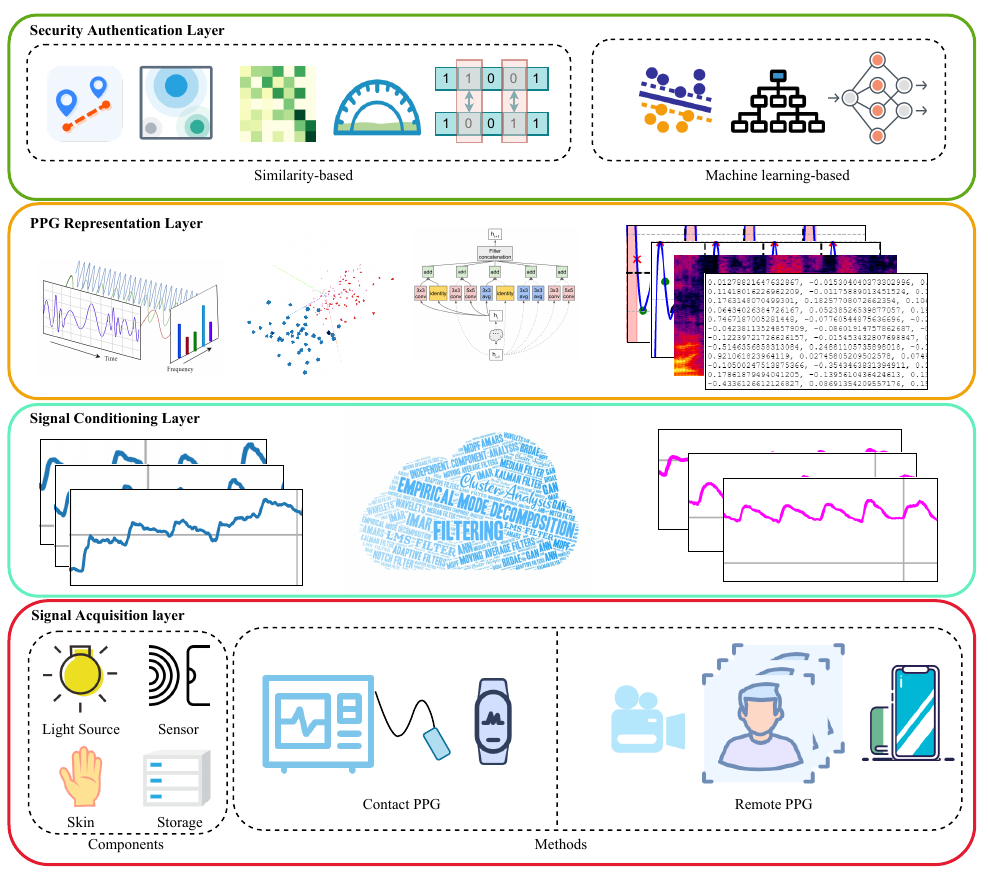}
  \caption{Four-layered PPG-based authentication framework. 
  Firstly, the PPG signal of the user can be captured using different devices.
  Then, the raw PPG signal is processed by signal conditioning to obtain a high-quality signal. 
  In the third layer, features are extracted from the processed signal. 
  Finally, each of these features is applied to different tasks according to their properties.}
  \label{workflow}
\end{figure*}

The PPG signals differed between individuals.
The signal can be affected by genetic and non-genetic factors, according to many PPG signal studies \citep{tegegne2020heritability,wang2021arrhythmia,PANAHI2021102863}.
Differences in PPG signals are observed between individuals, empowering the upgrade from pre-set passwords to PPG signals for user authentication. 
PPG signals were first applied in biometrics in 2003 \citep{Gu1222403}.
Subsequently, the derivatives of the PPG signal were used for biometric authentication \citep{Yao4353358}. 
The approach to individual feature matching has shifted from the initial calculation of the distance between features to deep learning classifiers \citep{RESITKAVSAOGLU20141}.

{
We attempt to comprehensively investigate PPG signals in authentication applications. 
PPG signals used in authentication systems can effectively capture users' cardiac dynamic behaviors \cite{gil2008discrimination}, which is not possible for traditional methods like fingerprints, iris, and many alike. 
We found the articles from Google Scholar, IEEEXplore, ACM Digital Library, ScienceDirect, and DBLP using various search terms --- ``PPG", ``photoplethysmogram", ``security", ``authentication", ``biometrics", and ``attack". 
We assessed the relevance of the articles to our investigation by examining their titles, abstracts, and keywords, ranging from the first PPG-based biometrics in 2003 to recently published articles in 2023. 
We kept the papers directly related to the intersection of PPG signals and cybersecurity applications. 
We prioritized articles with a substantial number of citations, indicating their influence and recognition within the research community. 
We focused on the articles published in top conferences and journals known for their rigorous review processes and wide readerships, such as the IEEE Symposium on Security and Privacy, the ACM Conference on Computer and Communications Security, Computers \& Security, IEEE Transactions on Information Forensics and Security and several others.
We paid attention to the paper authored by recognized experts or research groups in the field of cybersecurity. 
While we aimed to include recent research, we also considered foundational papers published in earlier years.}

A survey of heart biometrics was presented in \citep{Rathore3410158} for user authentication with heart signals, but it suffers from a primitive coverage in PPG signals with merely six papers. 
A review on wearable biometric systems was presented in \citep{Sundararajan3309550} with only a few acquisition methods for PPG signals.  
This paper aims to present a comprehensive review of the authentication method based on PPG signals.
The main contributions are summarized as follows:
\begin{itemize}
\item We systematically present PPG-based authentication associated with security threats.  
We propose a novel taxonomy to organize various systems from the technical and application perspectives to provide a comprehensive insight into PPG signals.

\item We survey the most recent research on PPG-based authentication from 2003 to 2023 and summarize the view to enable future researchers to apply the PPG signals technologies. 

\item We discuss the challenges of PPG-based authentication to highlight open issues for immediate attention and suggest possible countermeasures for future research.
\end{itemize}

The rest of this paper is organized as follows: 
We propose a four-layered view of PPG-based authentication in Section \ref{Methodology}. 
In Sections \ref{resistnoise}, \ref{representationconstruct} and \ref{authenticationmodel}, the literature review is presented on PPG-based authentication.
We review the usage of PPG signals in other authentication models in Section \ref{otherppg}.
Section \ref{future} discusses the challenges faced by PPG-based authentication and proposes the corresponding future directions. 
Section \ref{conclusion} concludes this paper.

\section{A Novel Four-Layered View on PPG-based Authentication}
\label{Methodology}

In this section, we present a novel view of PPG-based authentication. 
Fig.~\ref{workflow} presents our four-layered framework generalized from the literature. 
The bottom layer is the signal acquisition layer for collecting PPG signals. 
The second layer denoises the signal with the enhancement of its signal-to-noise ratio.
The third layer, called the PPG representation layer, extracts the signal's features through feature transformation and selection. 
The security application layer uses the extracted features for authentication. {Our framework was developed through meticulous information aggregation and generalization from diverse literature sources. 
We aim to capture and categorize the essential facets, factors, and dimensions prevalent in the existing body of knowledge. 
To provide further clarity, we emphasize that our taxonomy is not merely a subjective framework based on individual expertise. 
Instead, it is rooted in a systematic literature analysis, ensuring its relevance and coverage of the key elements within the field. 
By presenting this taxonomy, we contribute a structured and organized approach to the study of PPG signals in the context of cybersecurity, enabling researchers to navigate the complexities of this domain effectively.
}


\subsection{Signal Acquisition Layer}
The signal acquisition layer includes the actions for capturing the user's PPG signal.
It extracts PPG signals from the skin and converts them into electrical signals for transmission to the next layer.
This layer consists of four main components --- light source, skin, sensor, and storage. 
The blood flow in the skin is the source of the signal. 
The light source exposes the signal to the sensor. 
The sensor converts the received signal into an electrical signal to feed subsequent layers for processing. 
The mainstream sensors are photodetectors that convert the received light intensity into a voltage signal. 
A camera is regarded as a sensor for capturing rich information of light.
Storage determines the carrier of the signal, including electrical and video signals. 
Eventually, all signals are transformed into PPG waveforms and passed to the noise reduction layer.

Depending on the sensor and acquisition types, many methods are available to capture PPG signals. 
We classify them as contact and remote captures.  
The contact type captures the signal using photodetectors, and the device remains contacting with the skin. 
The remote type usually acquires the PPG signal by analyzing the video obtained by the camera, which allows the signal to be acquired at a certain distance.
Within these two types, there are also subtle differences in the different acquisition devices. 
We have compared four most common devices, including oximetry (contact), wearable devices (contact), smartphone cameras (remote), and HD cameras (remote). 
The oximeter and wearable devices capture reflected or projected light intensity changes primarily through light-sensitive sensors \citep{Daniel9112327,Sukhpreet9441342}. 
Smartphone cameras and HD cameras capture the change of RGB value among video frames to detect the change of blood flow in human skin tissue \citep{Aziz9385841,Yiming9358227}.  
Although the captured PPG signals all respond to a wealth of individual biometric information, the signal morphology acquired by various methods differs because tissues of different body parts emit different PPG signals.

{For a comprehensive comparison, we summarized five evaluation dimensions of signal acquisition from the existing literature. }

{
\begin{itemize}
\item  \textbf{Security} refers to the level of data protection and privacy provided during signal acquisition. 
It encompasses aspects such as encryption, authentication mechanisms, secure transmission protocols, and protection against unauthorized access. 
\item \textbf{Signal Quality} focuses on the acquired signals' accuracy, reliability, and fidelity. 
It involves evaluating factors such as noise levels, signal-to-noise ratio, resolution, dynamic range, frequency response, and any distortions or artifacts introduced during acquisition. 
\item \textbf{Cost} evaluation involves assessing the financial implications of different signal acquisition methods. 
It includes considerations such as the initial investment required for equipment, ongoing maintenance costs, licensing fees for software or algorithms, and any additional expenses associated with the acquisition process. 
\item \textbf{Range} examines the ability of a signal acquisition method to capture signals from a distance. 
It evaluates the acquisition system's range and effectiveness in scenarios where physical proximity to the signal source may be limited. 
\item \textbf{Mobility} refers to the portability, flexibility, and ease of use of a signal acquisition system. 
It considers factors such as device size, weight, power requirements, and the ability to deploy or move the system in various settings. 
\end{itemize}
}

\begin{figure}[t]
  \centering
  \includegraphics[width=\linewidth]{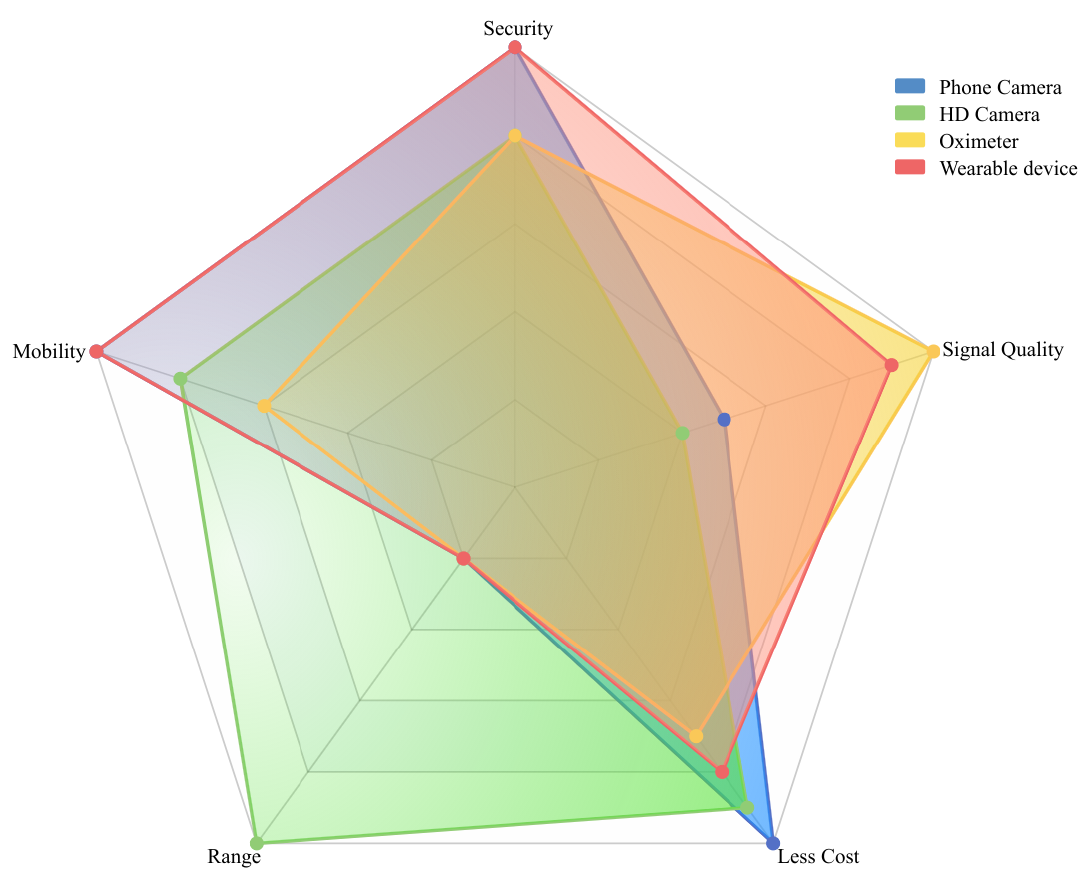}
  \caption{Characteristics of different acquisition methods. We compared the four most representative devices using the two acquisition methods on five dimensions. Regarding security, smartphones, and wearable devices performed the best. Regarding signal quality, the oximeter scored the highest. Phone cameras cost the least. HD cameras can capture PPG signals at a distance. Phone cameras and wearable devices have excellent mobility.}
  \label{radarpng}
\end{figure}

Fig.~\ref{radarpng} compares four acquisition methods in these five dimensions.
The pulse oximeter obtains high-quality signals partly because it isolates the interference from external ambient light.
However, a pulse oximeter needs to be clipped to a human finger, which interferes with any tasks that require finger involvement during continuous authentication. 
Due to the limited computational capability, oximeters transmit the captured signals to the endpoint for processing, increasing the risk of compromise. 
Wearable devices provide a new mode of interaction without affecting individuals' everyday lives, enabling continuous unnoticed authentication. 
A built-in physiological signal sensor allows wearable devices to capture PPG signals.
Unlike the traditional acquisition of PPG signals via photodetector, a phone camera acquires PPG signals by using the flashlight as the light source and shooting the fingertip on the camera \citep{Lovisotto9150630,ortiz2022biometric}. 
The HD camera method analyzes the face video for non-contact physiological measurements \citep{Patil8698552}, while the illumination usually comes from ambient light.
However, the PPG signals acquired using the camera are often low quality and noisy, especially for people with dark skin tones and quick motion artifacts due to body movement. 
In addition, the surrounding light conditions can significantly affect the signal quality. 
With the popularity of smartphones, HD cameras have been built into various devices, so this acquisition method incurs no extra cost.
For security reasons, the HD camera approach allows remote acquisition of PPG signals, which a malicious attacker can steal without the victim's awareness.

\subsection{Signal Conditioning Layer}
Noise is always present during any biomedical signal acquisition, no matter how well the devices are used  \citep{Mishra9298358}. 
Signal conditioning has become an important task for ensuring highly accurate authentication. 
The signal conditioning layer receives the raw PPG signal as the input and produces a high-quality PPG signal as the output.
Reducing or even eliminating noises in the signal is a primary concern when the types of noise need to be identified. 
The PPG signal contains rich heart-related information. 
Human bodies are usually assessed through statistical indicators (e.g., heartbeat interval, systolic peak) or physiological values (e.g., heartbeat rate, heart rate variability).
Hence, it is challenging to pinpoint the noise.

There are four primary types of noises: low-frequency noise, high-frequency noise, cardiac arrhythmia noise, and low-amplitude PPG signals. 
High-frequency and low-frequency noises are more commonly present in PPG signals than the other two. 
Specifically, motion artifacts (MA) are the most common low-frequency noise commonly found in wearable devices.
Both tissue deformation and sensor displacement may cause the appearance of motion artifacts \citep{nabavi2020robust}.
Another type of low-frequency noise is baseline wander noise.
Under normal circumstances, the centerline of the pulse wave signal is relatively smooth, indicating that the signal's non-pulsatile component is stable. 
However, the acquired signal has a constantly changing amplitude value of the overall waveform due to baseline wander caused by multiple factors, such as temperature variations, the bias of the instrumentation amplifier, and breathing motion \citep{Mishra9298358}. 
High-frequency noise is normally caused by power line interference, which refers to the ambient electromagnetic signal of the instrument amplifier and the power supply obstruction of the PPG recording probe. 
We can filter all the high-frequency and low-frequency signals directly by using the low-pass/high-pass filter at the cost of a significant loss of the original signal. 
Advanced filters like adaptive filter \citep{arunkumar2020casinor} help retain the maximum information of the original signal.

\subsection{PPG Representation Layer}
The representation layer receives the cleansed signal as the input before yielding feature vectors that apply to authentication systems.
Its primary objective is to extract features from the signal that are resilient to time and environmental changes while preserving the uniqueness of individual features.
The PPG representation layer comprises feature transformation and feature selection.
Fiducial points or statistical information can be directly extracted from the signal as feature vectors, like systolic peak, diastolic peak, and heart rate variability. 
The dicrotic notch is related to blood pressure \citep{mousavi2019blood}, and the systolic peak is associated with cardiovascular aging \citep{chiarelli2019data}.
Although these features can be acquired quickly from the raw signal, they are susceptible to changes in the surrounding environment and the physical state of the subject.

Feature transformation and feature selection are suitable for different tasks. 
Feature transformation converts the current feature space to a different space to acquire robust features for authentication, like from time-domain to frequency-domain. 
Feature selection helps remove redundant or irrelevant information. 
While removing the interference of useless information, feature selection also reduces features' dimensionality and computational cost.

\subsection{Security Application Layer}
The security application layer implements the authentication applications using features extracted from the PPG representation layer. 
PPG signals represent an individual's unique hemodynamic and cardiovascular system. Hence, PPG signals identify their owners during authentication.

The user authentication process comprises the enrollment phase and the authentication phase.
During the enrollment phase, the biometric system learns the feature vectors extracted from the individual. 
The enrollment phase can be regarded as the training phase from the machine learning perspective. 
The learned templates are stored on a local device or in the cloud as individual identifiers.
The authentication phase is further divided into two scenarios --- verification and identification.
Verification determines whether the user is consistent with the declared identity.
Identification attempts to find the best matching enrollment template in the system that corresponds to the user.
A biometric system can be regarded as a matching or classification problem.

At present, several methods distinguish the PPG signals of different individuals.
A straightforward method uses the similarity between features to distinguish the PPG signals between individuals.
A predefined threshold value determines the degree of similarity. 
If the similarity between features exceeds a preset threshold, the signals are considered to belong to the same individual. 
Distance and correlation are common approaches to measure similarity \citep{salanke2013enhanced, akhter2015heart, Yao4353358}.

User authentication is typically translated into a classification problem in machine learning as the paradigm where user profiles are associated with different classes. 
Features manually extracted through traditional machine learning do not guarantee an adequate representation of the uniqueness of individual PPG signals.
On the contrary, deep learning approaches are usually end-to-end solutions. 
Deep learning methods feed the training data and corresponding labels into the model before learning  useful features and inferring the testing set results. 
Deep learning methods are often preferred over manual feature extraction when we lack profound domain knowledge to understand the feature domain.


\section{Acquisition and Conditioning}
\label{resistnoise}

PPG signals consist of pulse signals as repetitive waveforms and motion artifacts as bursty signals.
Statistical differences (e.g., kurtosis, skewness, and standard deviation) can be applied to PPG signals for motion artifact detection \citep{Zhao3241539}.
According to the recoverability of cardiac signals, motion artifact is divided into two categories --- distal and proximal wrist \citep{Zhao9155526}. 
Distal wrist activity is a primary arm movement without involving the tendons and muscles of the wrist region. 
On the other hand, proximal wrist activities are horizontal and wrist-level movements that directly affect blood volume changes in the wrist region. 
Hence, proximal wrist activities may significantly impact PPG measurements from wearable devices.

Though distal wrist activity has a minor and recoverable effect on PPG signals, proximal wrist activity can have a long-lasting, intense, and non-recoverable effect on PPG measurements. 
Continuous near-wrist activity and accidental disease may cause sharp changes in heart conditions and affect the system's performance, resulting in a temporary reversion to a conventional authentication method like passwords.
When motion artifact is scattered or present in only a few contiguous segments, it is associated with distal wrist activity so that we can reconstruct the associated pulse waveform. 
When motion artifact is detected in consecutive PPG signals, the motion artifact occurrence is attributed to proximal wrist activity. 
Therefore, motion artifact removal helps eliminate the affected PPG segments.

Mobile phone cameras have become an easy choice to acquire PPG signals because mobile devices are widely popular \citep{Lovisotto9150630,ortiz2022biometric}. 
However, poor light conditions and frequent vibrations often affect the quality of PPG signals collected by mobile phone cameras.
Reliable cardiac motion patterns could only be obtained with the proper camera configuration and sufficient light entering the camera. 
Excessive (too low or too high) flashlight illumination reduces pixel sensitivity when capturing cardiac motion patterns from the camera. 
Thus, the camera configuration (i.e., flash intensity, ISO settings) needs adjustment to offset the variation of ambient light \citep{Liu3326093}. 
Dynamically selecting the pixels in the video captured by the camera, such as only a subset of the most sensitive pixels to heart motion or removing invalid pixel points, can improve the signal-to-noise ratio of heart measurements.

Since PPG sensors consist of LED and photodetector with specific spectral sensitivity and emission wavelengths, subtle differences in such devices are common. 
These signals collected from different devices can be considered data from different domains.
This problem can be handled by applying cross-domain adaptation methods \citep{Lee9053604}, like DRANet \citep{lee2021dranet} and PCS \citep{yue2021prototypical}. They are usually applied to vision-related tasks. 
It is possible to eliminate the non-pulsatile component of the signal by adding an amplifier bias adjustment circuit, obtaining a high signal-to-noise ratio pulsatile component from the original PPG signal \citep{Yongbo4406656}.
Improvements from a hardware perspective result in better signal quality and make identification data processing easier.

Additional factors affecting PPG signal quality are human body posture and emotions. 
If data were obtained while the participant was sitting steadily, the effects of physical exercises on PPG signals were often ignored. 
Significant differences in the PPG signals were observed among participants in the exercise state \citep{salanke2013enhanced}.
Besides exercises, the PPG signal reflects the influence of the autonomic nervous system on cardiac activity, which can easily be altered by changes in heart rate caused by mood fluctuations. 
Using a Gaussian function to represent the PPG signal features approximately has excellent robustness for emotions \citep{Sarkar7791193}. 
The classification of emotions in the datasets is based on participants' subjective perceptions.

As an authentication feature, feasibility is critical in long-term situations. 
The correlation coefficients of the PPG waveforms recorded during the month compared in \citep{Patil8698552} remain constant.
Because of the frequent acquisition during continuous authentication, the effect of time on the signal is not considered in \citep{Bonissi6656145}.
Empirically, the performance of the authentication model in the cross-session case declines over time \citep{sancho2017photoplethysmographic, Hwang9130730, hwang2021pbgan, Hwang9413906}. 
Feature selection helps identify features resilient to time \citep{Yadav8411233}.
Model fusion and generative adversarial networks improve the stability of the model over time \citep{hwang2021pbgan, Hwang9413906,liu2023dual,hwang2022new}.

\section{Representation Construction}
\label{representationconstruct}

Features representing PPG signals can be constructed in several different ways. 
Individual template vectors are built by extracting the number of peaks, time intervals, up slopes, and down slopes as features from a single-cycle PPG signal \citep{Gu1222403}.
In addition to these features, morphological features like the waveform area and the waveform angle were introduced in \citep{Lee7370629}. 
The features are obtained directly from the original waveform, implying potential interference of external factors like baseline wander and motion artifact.
This method of approximating the signal ignores the information of higher-order derivatives contained in the pulse.
Because the information contained in the PPG signals cannot be fully utilized to improve recognition accuracy and reliability, Yao \textit{et al.}~\citep{Yao4353358} proposed to consider both first- and second-order derivatives of the PPG signals.
The features obtained through higher-order derivatives are discriminative and sensitive to noise in the recognition task. 
In contrast, features from lower-order derivatives are more robust and less sensitive than those from their higher-order counterparts.

The feature transformation can obtain robust individual template vectors. 
Frequency-domain signals are generally more robust to time variations than time-domain signals.
The Fourier transform converts the signal from the time domain to the frequency domain \citep{Hwang9413906}.
However, the Fourier transform has an inherent flaw when dealing with non-smooth signals.
It only obtains information about which frequency components a segment of the signal contains instead of the exact moments when each component appears. 
Thus, two signals with different time domains may have the same spectrogram.  
In this case, the short-time Fourier transform can decompose the entire time domain of the signal into an infinite number of small processes of equal length \citep{donida2021biometric}. 
By setting the window length, we can obtain the frequency at a particular point in time. 
Nevertheless, it cannot meet the demand of the changing frequency of non-stationary signals, such as PPG signals. 
The components of various signals in nature at different frequencies have different time-varying characteristics. 
Generally, the spectral features of the lower frequency components change more slowly over time, while the higher frequency features change more rapidly. 
To obtain suitable frequency resolution and time resolution in different time-frequency regions, Patil \textit{et al.}~\citep{Patil8698552} used the Wavelet transform to decompose the signal across different time and frequency bands.
Mel-Frequency Cepstral Coefficients work on specific frequency components according to the nonlinear Mel scale \citep{siam2021ppg}.

To construct individual template vectors, feature selection improves the discriminability and robustness of the features.
Principal component analysis was used in \citep{Lovisotto9150630} to remove correlations between variables in a biometric system, retaining key features that effectively distinguish PPG signals from different individuals. 
However, principal component analysis can only perform linear transformations on the data, resulting in weak outcomes for linearly inseparable data. 
Hence, kernel principal component analysis is used in \citep{Zhang3297174} to map data that cannot be linearly classified in the low-dimensional space to the high-dimensional space for principal component analysis. 
In addition, various algorithms are used for feature selection in biometric systems, including linear discriminant analysis \citep{Yadav8411233} and genetic algorithm \citep{Karimian7953035}.

For instance, the waveform in a heartbeat cycle can be approximated by simple functions.
We can use some morphological modeling approaches to describe the PPG signals for biometrics quantitatively \citep{Cheng9074695}.  
Data need to be pre-processed before being manually extracted for features.
Conversely, deep learning automates the feature selection process that helps develop a fully data-driven end-to-end biometric system with PPG signals \citep{Luque8553585, Everson8350983}.

\section{PPG-based Authentication Model}
\label{authenticationmodel}

It is challenging to optimize, develop, or transform the training data structure to improve classification performance.
Among the similarity-based methods for identifying individual templates, the most common measure uses the Euclidean distance \citep{akhter2015heart, Gu1222403}.
Euclidean distance represents the straight line distance between two feature points in a Euclidean space. 
However, the Euclidean distance is susceptible to different feature scales in the vector. 
The Mahalanobis distance eliminates some limitations of the Euclidean metric, such as automatically considering the scaling of the axes, correcting for correlations between different features, and providing curved and linear decision boundaries \citep{salanke2013enhanced}.
The Mahalanobis distance calculates the covariance distance between two data points.
Pearson correlation is widely used to measure the degree of linear correlation between two variables \citep{Yao4353358}.
Among the above methods, a few outlier data in the training set can significantly affect the classification results because any similarity-based approach only needs to store a small number of training samples.

\begin{table*}[!ht]
\tiny 
\renewcommand\arraystretch{1.4}
\centering
\caption{Outline of reviewed papers attributes on user authentication. ``$\checkmark$": Will work. ``$\CIRCLE$": High level. ``$\LEFTcircle$": Medium level. ``$\Circle$": Low level. Permanence: The robustness of the authentication to temporal changes, including long time intervals and mood changes. {Time gaps within one day are evaluated as low level, while gaps ranging from one to seven days or mood changes are considered medium level. Gaps exceeding seven days are classified as high level.}  
Privacy: The potential exposure level of biometric signals. {For data acquisition methods, video analytics-based data collection is low level, photoelectric sensor-based methods are medium level, and integrating photoelectric sensors with authentication systems in the same device is high level.}
Cancelability: Whether the authentication template can be revoked/replaced. {The papers that incorporated cancellable  techniques have been marked. 
Wearability: The papers that have been marked signify the utilization of wearable devices.}
Transparency: If the user can perceive the authentication process. {They often require wearable devices or video-based analytics.}
Accessibility: Whether it is suitable for all populations, especially for people with physical disabilities. S: Single pulse. C: Continuous waveform. ``-": Not considered}
\setlength{\tabcolsep}{2.7mm}{
\begin{tabular}{lllllllllll}
\toprule
\textbf{Reference} & \textbf{Source} & \textbf{Assessment} & \textbf{Permanence} & \textbf{Privacy} & \textbf{Cancelability} & \textbf{Wearability} & \textbf{Continuity} & \textbf{Transparency} & \textbf{Accessibility} \\ \midrule

  \citet{Zhang10149364} & S        & Data-driven      & -          &   $\CIRCLE$     &     $\checkmark$           &       $\checkmark$           &     $\checkmark$        &        $\checkmark$        &    $\checkmark$           \\

  \citet{zhou2023gesture} & C        & Data-driven      & $\LEFTcircle$           &   $\CIRCLE$     &              &                &            &              &              \\ 
  
  \citet{liu2023dual} & C        & Data-driven      & $\CIRCLE$           &   $\LEFTcircle$       &              &                &            &   $\checkmark$              &  $\checkmark$   \\ 

   \citet{hwang2022new} & S        & Data-driven      & $\CIRCLE$          & $\CIRCLE$       &              &             &            &              &    $\checkmark$  \\ 

   \citet{wang2022campressid} & S        & Data-driven      & $\Circle$            & $\LEFTcircle$        &              &             &            &              &        $\checkmark$          \\

 \citet{kumar2022fusion} & C        & Data-driven      & $\Circle$           & $\CIRCLE$        &              &     $\checkmark$           &            &   $\checkmark$              &    $\checkmark$              \\ 

\citet{ortiz2022biometric} & S        & Data-driven      &  -            & $\LEFTcircle$        &              &             &            &              &    $\checkmark$              \\ 

\citet{ye2021ppg} & S        & Data-driven      & -         & $\CIRCLE$       &              &             &            &              &    $\checkmark$\\ 

\citet{hwang2021pbgan} & S        & Data-driven      & $\CIRCLE$          & $\CIRCLE$       &              &             &            &              &    $\checkmark$ \\ 

\citet{Hwang9413906} & S        & Data-driven      & $\CIRCLE$           & $\CIRCLE$       &              &             &            &              &   $\checkmark$ \\

\citet{siam2021ppg} & C & Data-driven      &     -       & $\CIRCLE$       &              &             &            &              &           $\checkmark$      \\ 
  
\citet{donida2021biometric} & C & Data-driven      &   -       & $\CIRCLE$       &              &             &            &              &    $\checkmark$  \\ 
  
 \citet{Hwang9130730} & S        & Data-driven      & $\CIRCLE$           & $\CIRCLE$       &              &             &            &              &    $\checkmark$   \\ 
  
\citet{Zhao9155526} & S        & Data-driven      &     -      & $\CIRCLE$       &              & $\checkmark$            & $\checkmark$           & $\checkmark$             &      $\checkmark$          \\ 
  
\citet{Lee9053604} & S        & Data-driven      & $\LEFTcircle$           & $\CIRCLE$       &              & $\checkmark$            &            & $\checkmark$             &      $\checkmark$          \\ 
  
\citet{Hinatsu9176311} & C & Data-driven      &    -        & $\CIRCLE$       &              &             &            &              &     $\checkmark$  \\
  
\citet{Cao9155380}                   & S        & Data-driven      & $\CIRCLE$             & $\CIRCLE$       & $\checkmark$             & $\checkmark$            &            &              &      $\checkmark$            \\

\citet{Lovisotto9150630} & S        & Data-driven      & $\LEFTcircle$             & $\LEFTcircle$        &              &             &            &              &    $\checkmark$              \\ 
  
\citet{Biswas8607019} & C & Data-driven      &  -         & $\CIRCLE$       &              & $\checkmark$            &            & $\checkmark$             &      $\checkmark$            \\

\citet{Hwang9052394} & C & Data-driven      &   -       & $\CIRCLE$       &              &             &            &              &    $\checkmark$ \\ 
  
\citet{Liu3326093} & S        & Data-driven      & $\CIRCLE$            & $\LEFTcircle$        &              &             &            &              &        $\checkmark$          \\ 
  
\citet{Cheng9074695} & S        & Data-driven      &   -       & $\CIRCLE$       &              &             &            &              &     $\checkmark$  \\ 
  
\citet{Shang8802738}  & C & Data-driven      &       -     & $\CIRCLE$       &              & $\checkmark$            &            & $\checkmark$             &    $\checkmark$             \\ 
 
\citet{Zhang3297174} & S        & Data-driven      &    -       & $\CIRCLE$       &              &             &            &              &   $\checkmark$    \\ 
  
\citet{Everson8350983} & S        & Data-driven      &   -        & $\CIRCLE$       &              & $\checkmark$            &            & $\checkmark$             &       $\checkmark$         \\ 
  
\citet{Luque8553585}  & C & Data-driven      &   -        & $\CIRCLE$      &              &             &            &              &         $\checkmark$      \\

\citet{Yadav8411233} & S        & Data-driven      & $\LEFTcircle$           & $\CIRCLE$       &              &             &            &              &   $\checkmark$   \\
  
\citet{Zhao3241539} & S        & Data-driven      &   -      & $\CIRCLE$       &              & $\checkmark$            & $\checkmark$           & $\checkmark$             &      $\checkmark$         \\ 
 
\citet{Patil8698552} & C & Data-driven      & $\LEFTcircle$        & $\Circle $        &              &             &            & $\checkmark$             & $\checkmark$              \\
  
\citet{Karimian7953035} & S        & Data-driven      &   -         & $\CIRCLE$       &              &             &            &              &  $\checkmark$   \\ 
  
\citet{sancho2017photoplethysmographic}  & S        & Statistics-based & $\LEFTcircle$            & $\CIRCLE$        &              &             &            &              &   $\checkmark$  \\ 
  
\citet{Ohtsuki7794969} & C & Data-driven      &    -       & $\CIRCLE$       &              & $\checkmark$            &            &              &                \\
  
  
\citet{Jindal7592193} & S        & Data-driven      &     -      & $\CIRCLE$       &             &     $\checkmark$         &            &     $\checkmark$          &     $\checkmark$          \\ 
  
\citet{Sarkar7791193} & S        & Data-driven      & $\CIRCLE$             & $\CIRCLE$       &              &             &            &              &     $\checkmark$  \\

\citet{akhter2015heart} & C & Statistics-based & $\Circle $            & $\LEFTcircle$      &              &             &            &              &      $\checkmark$          \\ 
  
\citet{Lee7370629}  & S        & Data-driven      &     -      & $\CIRCLE$       &              &             &            &              &    $\checkmark$  \\
 
\citet{RESITKAVSAOGLU20141} & S        & Data-driven      &     -      & $\CIRCLE$       &              &             &            &              &    $\checkmark$      \\ 
  
\citet{Bonissi6656145} & S        & Statistics-based &   -       & $\CIRCLE$       &              &             & $\checkmark$           &              &       $\checkmark$        \\ 
  
\citet{salanke2013enhanced} & S        & Statistics-based &    -      & $\CIRCLE$       &              &             &            &              &   $\checkmark$       \\ 
 
\citet{Spachos6004938} & S        & Data-driven      &    -       & $\CIRCLE$       &              &             &            &              &     $\checkmark$     \\ 
  
\citet{Yao4353358}  & S        & Statistics-based &     -    & $\CIRCLE$       &              &             &            &              &    $\checkmark$        \\  
\citet{Gu1222403} & S        & Statistics-based &      -    & $\CIRCLE$      &              &             &            &              &       $\checkmark$         \\ \bottomrule
\end{tabular}
}

\label{tab_outline_userauthentication}
\end{table*}

Eight machine learning-based methods are often used for user authentication based on PPG signals: linear discriminant analysis \citep{Sarkar7791193}, support vector machine \citep{donida2021biometric, Zhao9155526, Hinatsu9176311,Lovisotto9150630, Karimian7953035, Zhang3297174}, k-nearest neighbor \citep{RESITKAVSAOGLU20141, donida2021biometric}, random forest \citep{Hinatsu9176311,Cheng9074695, Ohtsuki7794969, Cao9155380}, gradient boosted trees \citep{Lovisotto9150630, Zhao3241539, Zhao9155526}, multi-layer perceptron \citep{Karimian7953035,siam2021ppg}, restricted Boltzmann machines \citep{Jindal7592193}. 
These models are usually trained using the features from the feature capture layer as input and the class of the user as output.

Convolutional neural networks (CNNs) are popular for their wide range of applications in computer vision-related tasks. 
Recently, PPG-based user authentication has applied a CNN model \citep{Luque8553585}.
A typical CNN architecture consists of a convolutional layer, a pooling layer, and a fully connected layer. 
The target's low-level (points in the signal) and high-level features (overall trend of the signal) can be extracted by stacking the convolutional layers. 
Pooling layers are sampled to reduce the feature space while retaining the important features. 
The primary role of the fully connected layer is to classify the signal based on the features previously extracted from the convolutional and pooling layers.
In a CNN, the signal from each neural network layer propagates up one layer, and the samples are processed independently each time.

However, the PPG signals are time-series data, and the information on the time dimension is valuable.
LSTM adds a gate mechanism and a memory unit to Recurrent Neural Network (RNN)  to capture the long-term dependence of the input sequence by recording information from different periods. 
Therefore, the LSTM component captures long-time contextual information \citep{Everson8350983, Hwang9052394,Biswas8607019,Hwang9130730,ye2021ppg}. 
It also solves the gradient disappearance and gradient explosion problems in RNN. 
Many solutions like the transformer model \citep{vaswani2017attention} learn from sequence data. 
The current research on deep learning models in PPG-based authentication is limited and requires further exploration.

Biometric systems based on a single PPG signal are vulnerable since the acquisition equipment, and recording environment has a significant impact on the performance of the system. 
A PPG signal collected with a precise sensor in a controlled environment is reliable. 
However, if the PPG signal is unstable, an additional biometric signal can improve the result \citep{Spachos6004938}. 
ECG can be recorded simultaneously with PPG and provide a multi-fact biometric system.
The sensor can acquire the ECG and PPG signals simultaneously, thus synchronizing the ECG and PPG values. 
The systolic peak of PPG and the R-peak of ECG can be used to obtain the Pulse Transit Time and Pulse Arrival Time to match the user template, detecting any spoofing signal \citep{Karimian9086770}. 
To bypass the anti-spoofing system, attackers need to measure the victim's ECG and PPG at the same time. 
Even if the attacker is able to generate the victim's ECG and PPG, matching them from the same time domain would be challenging. 
ECG signals require the user to use additional measurement equipment, increasing the system's complexity.
Ultra-wideband radar can measure the user's breathing pattern and synchronize with PPG signals so that it can be used to detect unknown presentation attacks \citep{Forouzanfar9521212}.
Moreover, fusion-ID authenticates users by fusing PPG signals with information from motion sensors \citep{kumar2022fusion}.

Table~\ref{tab_outline_userauthentication} summarizes the concept of the user authentication-related articles we reviewed. Most studies use a single heartbeat cycle of the PPG signal as a unique identifier, as it is easier to extract individually relevant information. 
{Permanence pertains to the ability of an authentication system to accurately identify and authenticate individuals over time, despite variations that may occur due to the passage of time or changes in an individual's mood. 
It implies that the system can effectively recognize and verify an individual's identity consistently, regardless of time gaps between authentication attempts or fluctuations in their emotional state. 
In Permanence, the evaluation of time gaps within one day is considered low level, and between one and seven days are considered medium level, longer than seven days are considered high level. 
Privacy concerns arise in PPG-based user authentication methods due to collecting and storing sensitive biometric data, specifically pulse or blood flow patterns. 
Privacy concerns also involve evaluating the potential risks of unauthorized access or data breaches associated with the methods. 
For example, video analytic-based data collection methods pose a higher risk of data leakage than traditional photoelectric sensor-based data collection methods. }

We also found that there is no standard to evaluate PPG-based authentication. 
Table.~\ref{tab_outline_userauthentication} summarizes five evaluation metrics (Cancelability, Wearability, Continuity, Transparency, and Accessibility). 
To improve the practicality of PPG-based user authentication, further research is needed in these five aspects.

\textbf{Cancelability:} Biometric systems usually require biometrics to be permanent. 
However, once the biometric template is exposed, the threat to the identification system is permanent. 
Cancelability means that the template can be replaced in biometric template exposure.
{The raw biometric data undergo a non-invertible transformation creating a new biometric template. 
This transformation could be unique for each application, providing protection across systems. 
If a system is compromised and the biometric templates are stolen, these templates cannot be used, and a new transformation can be applied to generate new templates, essentially canceling the old ones.}
The most straightforward revocable authentication is to encrypt the biometrics in the device.
In PPG-based user authentication, feature transformations are used to map features into different vector spaces to cancel templates. 
{Cancelability can be quantified by two main aspects --- revocability and unlinkability \cite{bedari2021design}. 
Revocability ensures that the newly generated one will not reduce the authentication performance when a biometric template is compromised. 
Unlinkability refers to the inability to establish a link between the original biometric features and the newly generated ones. 
If such a link is identifiable, it might be possible to recreate the original biometric data from the new features, defeating the revocation purpose. 
As listed in Table \ref{tab_outline_userauthentication}, the papers that incorporated cancellable techniques have been marked. 
We can find that most of the papers ignore the assessment of cancelability.
}

\textbf{Wearability:} 
{Wearability refers to the suitability and practicality of incorporating biometric sensors or devices into wearable technology or accessories. 
This concept emphasizes the ability of these devices to comfortably and unobtrusively collect and analyze biometric data from individuals in their everyday activities. 
The goal is to provide seamless and continuous biometric authentication or monitoring while ensuring user comfort, convenience. }
With the miniaturization of physiological signal sensors, most wearable devices have these sensors built-in for healthcare. 
For wearable authentication, PPG signals are primarily collected by wristband devices.
These wristband devices are easily accessible and usually inexpensive. 
{In Table~\ref{tab_outline_userauthentication}, the papers that have been marked signify the utilization of wearable devices (e.g., smartwatches and wristbands) for signal acquisition. }

\textbf{Continuity:} Authentication is usually performed only on the first access in most authentication scenarios.
The user identity is maintained by the credentials obtained through authentication.
It may lead to security risks for subsequent operations.
For example, if a legitimate user leaves the device unattended, a malicious user accessing the device will potentially access other services.
Continuous authentication enables continuous verification of the user's identity for the entire duration of the session.
While traditional continuous authentication methods typically rely on transient events, PPG signals are continuous waveforms that can easily provide non-intrusive continuous authentication. 
{We have marked the papers that reported the continuous authentication performance of their methods in Table~\ref{tab_outline_userauthentication}. }

\textbf{Transparency:} 
{Transparent authentication refers to an authentication process that is seamless, unobtrusive, and user-friendly. 
It aims to provide a frictionless user experience by minimizing user intervention or explicit authentication actions. 
In transparent authentication, the user's identity is verified in the background or implicitly through various methods or factors without requiring explicit input.}
Wearable device-based PPG user authentication offers the possibility of transparent user authentication. 
It reduces the probability of a spoofing attack since the user does not know when the authentication occurred. 
{In Table~\ref{tab_outline_userauthentication}, the Transparency column excludes methods that necessitate active user participation. }

\textbf{Accessibility:} 
{It refers to the authentication methods and practices designed to accommodate individuals with disabilities or impairments. 
It aims to ensure that individuals with diverse abilities can access and utilize digital systems securely and conveniently.
In the context of accessibility authentication, traditional authentication methods may present barriers for individuals with disabilities. 
For example, individuals with visual impairments may encounter difficulties in entering complex passwords or reading visual authentication cues, while those with motor impairments may struggle with physical interactions like typing or using traditional input devices.
PPG signals can be collected in multiple body parts like ears, forehead, fingers, and toes, implying high accessibility. 
From Table~\ref{tab_outline_userauthentication}, it can be observed that all methods listed are considered accessible, except for those that necessitate gestural involvement.
}

\section{Miscellaneous Authentication Models with PPG Signals}
\label{otherppg}
Though face recognition is the most widely used biometric feature, current face recognition systems are vulnerable to spoofing attacks. 
Face recognition systems may fail in front of highly realistic 3D masks because they capture local facial details to distinguish real faces from fake ones.
Because PPG signals are present only in natural living tissue and absent in surface materials of any mask or printed material, facial liveness can be detected by finding PPG signals in facial videos \citep{Chen8057220}.
Remote photoplethysmogram (rPPG) signals are present in an organic face, resulting in the color value of facial areas in the video varying with the heart pulse. 
Hence, the peak amplitude of the rPPG spectrum could reflect the heartbeat intensity.
The observed amplitude is susceptible to environmental noises due to illumination and camera settings. 
Moreover, the noise may dominate the observed signal.
Cross-correlation operations of local rPPG signals at different face regions to amplify the shared heartbeat frequency can suppress the interference of non-periodic noise \citep{Liu0_34}.

DeepFake \citep{li2020celeb} uses a generative adversarial network to forge a face to replace the original face in the video clip.
DeepFake poses a real threat to the accuracy of the multimedia information available, especially since falsifying a politician's speech may lead to harmful results.  
Live detection for face recognition mainly relies on detecting heart rate, while heart rate may be present in a DeepFake video clip with a slightly different pattern of PPG signals.
Videos generated by DeepFake can be identified by how consistent the regular heart rate in the facial area is  \citep{qi3413707}.

Handwritten signature authentication prevents fraud in financial, judicial, and administrative settings. 
Traditional handwritten signature authentication requires historical samples because it only compares static handwriting with the user's previous handwriting to determine the signature's authenticity.
Several methods have been used to automatically generate models for spoofing handwritten signature images \citep{rahman2022ppgsign,li2021black}. 
PPGSign \citep{hafemann2019characterizing} uses the PPG signal collected from a wrist-worn wearable device to verify a user's handwritten signature.
Unlike traditional PPG-based authentication, PPGSign studies the dynamic component of the PPG signals caused by hand movements. Moreover, gestures can be used to assist in authentication by changing the signal shape \cite{zhou2023gesture}.

\section{Research Gaps and Future Work}
\label{future}
\renewcommand\arraystretch{1.3}
\begin{table*}[!ht]
\scriptsize 
\centering

\caption{Comparison of publicly available PPG datasets in different dimensions. Subjects: Number of participants in the dataset. Location: Which body part the signal was collected. F: Face. FT: Fingertip. W: Wrist. E: Ear. FH: Forehead. ``$\CIRCLE$": Cross-session.  ``$\Circle$": Single-session. Patient: The participant is under medical supervision. Relax: Participants remain as stationary as possible during signal acquisition. Exercise: Including Running, Cycling, Walking, and Climbing. Emotion: Use games or videos to stimulate participants' emotions. ``-": The health status of the participants was not considered. ``$X$": There were health problems among the participants. ``$\checkmark$": All participants are healthy. }
\begin{tabular}{lllllllll}
\toprule
\textbf{Reference}& \textbf{Subjects}    & \textbf{Location} & \textbf{Time Span} & \textbf{Status} & \textbf{Storage}  &  \textbf{Acquisition Environment} & \textbf{Health Status}    \\ 

\midrule

 \begin{tabular}[c]{@{}l@{}}UBFC-Phys\footnotemark[1] \\ \citep{sabour9346017} \end{tabular}
 & 56 & F & $\Circle$ & Exercise  & Video  & Laboratory & $\checkmark$ \\

\begin{tabular}[c]{@{}l@{}}Biosec3 \\ \citep{Hwang9413906} \end{tabular}
  & 170  & FT & $\CIRCLE$  & Exercise  & Electrical signals  & Office & $\checkmark$ \\

 \begin{tabular}[c]{@{}l@{}}SeeingRed \\ \citep{Lovisotto9150630} \end{tabular}
  & 15  & FT& $\CIRCLE$ & Relax & Video  & Laboratory & - \\

 \begin{tabular}[c]{@{}l@{}}PPG-ACC \\ \citep{biagetti2020dataset} \end{tabular}  & 7   & W & $\Circle$ & Exercise & Electrical signals  & Laboratory & $\checkmark$  \\

 \begin{tabular}[c]{@{}l@{}}UBFC-RPPG \\ \citep{bobbia2019unsupervised} \end{tabular}
 & 50 & F & $\Circle$ & Exercise  & Video  & Laboratory & $-$ \\

\begin{tabular}[c]{@{}l@{}}TokyoTech \\ \citep{Maki8857081} \end{tabular}
 & 9 & F & $\Circle$ & Relax  & Video & Laboratory & $-$ \\

\begin{tabular}[c]{@{}l@{}}CIME-PPG \\ \citep{xu2019photoplethysmography} \end{tabular}
 & 48 & FT & $\Circle$ & Exercise  & Electrical signals  & Laboratory & $X$  \\ 

\begin{tabular}[c]{@{}l@{}}PPG-DaLiA \\ \citep{reiss2019deep} \end{tabular}
& 15 & W & $\CIRCLE$  & Exercise  & Electrical signals  & Wild & $\checkmark$   \\

\begin{tabular}[c]{@{}l@{}}GYRO-ACC \\ \citep{Lee8529266} \end{tabular}
 & 24 & W & $\Circle$  & Exercise  & Electrical signals  & Laboratory & $\checkmark$   \\

 \begin{tabular}[c]{@{}l@{}}VIPL-HR \\ \citep{niu2018vipl} \end{tabular}
  & 42 & F & $\Circle$ & Exercise  & Video & Laboratory & $-$ \\

\begin{tabular}[c]{@{}l@{}}OBF \\ \citep{li2018obf} \end{tabular}
 & 106 & F & $\Circle$  & Exercise  & Video & Laboratory & $X$    \\

\begin{tabular}[c]{@{}l@{}}LGI-PPGI \\ \citep{pilz2018local} \end{tabular}
  & 25 & F & $\Circle$  & Exercise  & Video & Wild & $-$    \\

\begin{tabular}[c]{@{}l@{}}PPG-BP\\ \citep{liang2018new} \end{tabular}
 & 219 & FT & $\Circle$  & Relax & Electrical signals  & Clinical & $X$   \\ 

\begin{tabular}[c]{@{}l@{}}PulseID \\ \citep{Luque8553585} \end{tabular}
 & 43 & FT & $\Circle$  & Relax & Electrical signals  & Office & $\checkmark$    \\ 
 
 \begin{tabular}[c]{@{}l@{}}Biosec1\footnotemark[2] \\ \citep{Yadav8411233} \end{tabular}
  & 41 & FT & $\CIRCLE$   & Exercise & Electrical signals  & Office & $\checkmark$    \\

 \begin{tabular}[c]{@{}l@{}}COHFACE\\ \citep{heusch2017reproducible} \end{tabular}
  & 40 & F & $\Circle$   & Relax & Video & Laboratory & $\checkmark$    \\

  \begin{tabular}[c]{@{}l@{}}Vortal\footnotemark[3]\\ \citep{charlton2016assessment} \end{tabular}
 & 57 & FT/E & $\Circle$   & Relax & Electrical signals   & Laboratory & $\checkmark$     \\ 

  \begin{tabular}[c]{@{}l@{}}MIMIC-III\footnotemark[4] \\ \citep{johnson2016mimic} \end{tabular}
 & 10,282 & FT & $\CIRCLE$    & Patient & Electrical signals   & Clinical & $X$     \\

   \begin{tabular}[c]{@{}l@{}}TROIKA\\ \citep{Zhang6905737} \end{tabular}
& 20 & W & $\Circle$   & Exercise & Electrical signals   & Laboratory & $-$     \\ 

    \begin{tabular}[c]{@{}l@{}}PURE\\ \citep{stricker2014non} \end{tabular}
 & 10 & F & $\Circle$   & Exercise & Video   & Laboratory & $-$     \\

\begin{tabular}[c]{@{}l@{}}TBME \\ \citep{karlen2013multiparameter} \end{tabular}
   & 42   & FT & $\Circle$  & Patient & Electrical signals  & Clinical & $X$ \\

    \begin{tabular}[c]{@{}l@{}}DEAP\\ \citep{koelstra2011deap} \end{tabular}
  & 32 & FT & $\Circle$   & Emotion & Electrical signals   & Laboratory & $\checkmark$     \\ \bottomrule
\end{tabular}

\label{tab_datasets}
\end{table*}

Many studies propose to use PPG signals for authentication because PPG signals have unparalleled advantages over traditional biometric features.
However, research on PPG-based authentication is in its infancy, especially when interacting with artificial intelligent models. 
To help future research, we discuss the current challenges and future research directions.

\subsection{Challenges in User Authentication}
The first challenge for PPG-based user authentication is \textbf{signal quality}.  
As PPG is a physiological signal, PPG signals' quality is subject to persistent changes under various factors. 
The variation may exaggerate potential vulnerabilities of the authentication application.
The signal quality may be affected in the following two aspects:

\begin{itemize}
 \item[] \textbf{The influence of intrinsic factors}: PPG changes over time, implying the necessity to consider single or multiple authentication sessions. 
Most existing studies investigate the single session when continuous signals are measured simultaneously. 
However, in practical applications, many scenarios are cross-session when the enrollment and authentication phases occur across different sessions \citep{Hwang9052394,Lovisotto9150630,sancho2017photoplethysmographic}.

The performance of cross-sessions in authentication results is worse than that of single session \citep{Hwang9130730, Hwang9413906}. 
It indicates that the current approach is not robust to the change of PPG signals as time varies. 
Furthermore, human emotional changes significantly impact the PPG signals. 
The influence of emotions in certain situations can help resist unauthorized certifications like enforcing a convict to authenticate. 
When the user is anxious to authenticate, the influence of emotions is counter-productive. 
In studies of the effect of emotion on PPG signals, watching a video or playing a game is investigated to stimulate participants' emotions. 
However, watching videos and playing games introduce many uncontrollable parameters, resulting in unreproducible results and conclusions. 
We cannot objectively determine their true emotions through the participants' descriptions, so significant misinformation may be present in the collected data.


\item[] \textbf{The influence of external factors}: 
External factors that affect PPG signals include light conditions, physical movement, skin temperature, and skin tones. 
PPG signals are collected by following the optical principle, implying that the external lighting conditions affect the signal quality. 
Wearable devices are a popular choice for capturing PPG signals, but the collected PPG signals are often affected by motion artifact noises caused by the physical movement of the wearer. 
Moreover, skin temperature and skin tone affect the quality of the PPG signal. 
\end{itemize}

The second challenge is the availability of high-quality \textbf{dataset}. 
Table~\ref{tab_datasets} {compares the publicly available datasets, focusing on the common features. 
These metrics were chosen according to their widespread use in the literature, their relevance to our research objectives, and their ability to provide a holistic understanding of the dataset characteristics.  
The information presented in the table is derived solely from the dataset descriptions.}
The most extensive dataset with different states has merely 170 participants' signals. 
It is challenging to collect an extensive data set in different states (movement status and emotions) as a physiological signal. 
Moreover, the interval between their measurements was only 18 days. 
Most existing datasets consider PPG signals collected in the resting state. 
The controlled environment in the experiment is different from our daily life, indicating that the signal noise in the data is significantly less than that in the real-world application.


The third challenge is the \textbf{overhead of the device}, especially in continuous authentication. 
Continuous authentication requires sensors to continuously monitor the user's physiological signals, implying the need for additional computational resources and energy overhead.
These overheads are significant issues for resource-limited wearable devices and smartphones.


Moreover, \textbf{data leakage} is another challenge. 
Though PPG signals are not easily leaked, the leaked PPG signals will threaten the security of the authentication system once the leak occurs. 
Furthermore, the development of radar and remote PPG to collect heart rate information makes it impossible to ignore the potentially severe consequences of data leakage.
Current research about cancelability focuses on the cancelable template. 
When a user template is compromised, it is replaced by redeploying a new one. 
However, it does not consider when the raw signal is leaked.
In addition, the wearability and transparency of authentication require the support of wearable devices.
All the authentication system components will be exposed to the adversary for stolen wearable devices.



Most existing work investigates medical-grade devices. 
With the popularity of wearable devices and the development of video technology, we believe that PPG signal-based security technology will be further developed in the future. 
Other physiological signals also receive increasing attention as promising candidates for implicit authentication systems. 
Some recent publications introduce biometric applications for authentication, including brain biometrics \citep{arias2021inexpensive}, ECG signals  \citep{hosseinzadeh2021electrocardiogram}, and electrical muscle stimulation \citep{Chen3445441}.

\footnotetext[1]{\url{https://ieee-dataport.org/open-access/ubfc-phys-2}}
\footnotetext[2]{\url{https://www.comm.utoronto.ca/~biometrics/PPG_Dataset/}}
\footnotetext[3]{\url{https://peterhcharlton.github.io/RRest/vortal_dataset.html}}
\footnotetext[4]{\url{https://physionet.org/content/mimiciii/1.4/}}

\subsection{Attack Threats}
Although it is challenging to steal unobservable PPG signals, PPG-based authentication faces potential threats. 
Two main types of attack threats are stealing user templates through leaked signals and attacks against user authentication AI models.

\noindent
\textbf{Stealing user templates}: With the development of biomedi-cine, many studies show that contactless methods can be used to detect heartbeat signals \citep{dasari2021evaluation}. HD cameras-collected rPPG signal is a severe threat to the PPG-based security system because of its easy-to-acquire and long-distance-use characteristics. 
The rPPG signal can acquire 70\% of the IPI information obtained by the contact sensor \citep{Calleja3}. 
When using rPPG to estimate IPI, darker skin has a higher average bit error rate, and it is more challenging to detect IPI accurately. 
This is due to the higher melanin content in darker skin than in lighter skin, reducing the diffuse reflection containing pulsation information, thus reducing signal quality. 
The head rotation also affects the accuracy of rPPG because it changes the light reflected from the skin.
In addition, compression of the video causes signal artifacts that can lead to false detection of heartbeats, but it does not significantly affect the detection of IPI.  
Although rPPG has been successfully applied to detect 3D mask presentation attacks and DeepFake videos, it is susceptible to environmental noise due to the particularly weak signal. 
rPPG is often used to obtain simple time- and frequency-domain features such as HRV and IPI to attack the corresponding security systems.  
The camera is susceptible to the user's background environment as the victim's environment changes in real-world scenarios.

Another non-contact method of detecting heartbeat signals is based on ultra-wideband radar. It measures the heartbeat by the variation in the amplitude and the arrival time of the reflected pulses. 
PPG is an optical signal that cannot be detected directly by radar. 
This setup allows radar-based methods to detect only heart rate information such as HRV and IPI of the heartbeat. 
Therefore, it is used in the same way as the HD camera approach, mainly for attacking systems based on simple features such as HRV and IPI. 
It does not mean that the HD camera- and radar-based approach is not a threat. 
There is already research to obtain high-quality rPPG signals using generative models \citep{lu2021dual}. 
Furthermore, radar information for reconstructing the ground truth PPG signal is also a possible threat. 
\citep{Yamamoto9139941} hypothesized the potential to reconstruct the ECG signal using the information collected by the doppler sensor. 
However, there is no research evidence using the doppler sensor to reconstruct PPG signals.

Once the PPG signal is compromised, it can be simulated using dynamic models. 
Gaussian functions can be applied to construct the mapping function that converts the attacker's PPG signals into dynamic model parameters similar to those of the victim to deceive the biometric system. 
We refer to this attack method as a gray-box evasion attack. 
The gray-box evasion attack only attracts limited attention due to its strong assumption of obtaining the victim's PPG signal in advance.

\noindent
\textbf{Attacking user authentication AI models}: Currently, there are many attack methods against machine learning that have tremendous potential \citep{chen2021real,lovisotto2020biometric}. For instance, \citep{chen2021real} spoofs speaker recognition systems by generating adversarial examples.
Adversarial examples refer to the addition of imperceptible perturbations to the original input to mislead the model and produce incorrect outputs. 
To the best of our knowledge, there are no defenses against PPG-based authentication adversarial examples. 
Traditional adversarial defenses are usually divided into two categories, detecting adversarial examples and improving the robustness of the classifier to adversarial examples (e.g., adversarial retraining and distillation). 
However, even with the state-of-the-art defense approach, there are still effective  attacks \citep{rosenberg2021adversarial}. 
Poisoning attacks on the model were performed in \citep{lovisotto2020biometric} through the update process of unsupervised templates. 
Since biometric systems usually adopt a self-renewal strategy, they are prone to poisoning attacks. 
Another attack that targets user authentication AI models is the backdoor attack. 
Inserting backdoors into the model makes the model trigger different results when faced with a specific symbol \citep{wang2019neural}. 
Unlike poisoning attacks, backdoor attacks can be hidden until the input activates them.
Although the backdoor attacks can be mitigated by pruning neurons \citep{shokri2020bypassing}, the mitigation is limited, and further exploration of possible measures remains future work. 
Each of these approaches is a potential threat to machine learning-based biometric systems.

\section{Conclusion}
\label{conclusion}
Traditional biometric authentication is susceptible to the threat of presentation attacks. 
Physiological signal-based authentication has recently received much attention, especially PPG signals.
PPG-based authentication becomes popular because of its non-intrusiveness, capability of continuous monitoring, spoof defection, and wide availability. 
This paper surveys PPG-based authentication in three aspects --- signal extraction, signal conditioning, and feature conversion and selection. 
The existing research review identifies the challenges, and future directions are proposed to match the various limitations. 
In addition, the attack threats against PPG-based authentication are summarized.
Thus, this survey can help researchers understand PPG signal-based applications' current development in security and future research trends.
Most studies in this review were conducted within the last few years, indicating a fast-growing interest in applying PPG signals among researchers in the security community. 
This paper shows the broad potential of using PPG signals for authentication.









\printcredits

\bibliographystyle{cas-model2-names}

\bibliography{COSE}

\bio{}
\endbio

\endbio

\end{document}